\documentclass[aps,showpacs,prd,twocolumn]{revtex4}%
\usepackage{graphicx}
\usepackage{amssymb}
\usepackage{amsmath}
\usepackage{color}
\usepackage{float}
\usepackage{ulem}
\usepackage{accents}
\usepackage{graphicx}
\usepackage{graphicx}
\usepackage{amsfonts}
\usepackage[colorlinks=true,pdfstartview=FitV,linkcolor=blue,citecolor=blue,urlcolor=blue,breaklinks=true]%
{hyperref}%
\providecommand{\U}[1]{\protect\rule{.1in}{.1in}}
\begin{document}
\title{\textbf{General CPT-even dimension-five nonminimal couplings between fermions
and photons yielding EDM and MDM}}
\author{Jonas B. Araujo}
\email{jonas.araujo88@gmail.com}
\author{Rodolfo Casana}
\email{rodolfo.casana@gmail.com}
\author{Manoel M. Ferreira Jr}
\email{manojr.ufma@gmail.com}
\affiliation{Departamento de F\'{\i}sica, Universidade Federal do Maranh\~{a}o, Campus
Universit\'{a}rio do Bacanga, S\~{a}o Lu\'{\i}s - MA, 65080-805 - Brazil}

\begin{abstract}
In this letter, we examine a new class of CPT-even nonminimal interactions,
between fermions and photons, deprived of higher order derivatives, that yields
electric dipole moment (EDM) and magnetic dipole moment (MDM) in the context
of the Dirac equation. The couplings are dimension-five CPT-even and
Lorentz-violating nonminimal structures, composed of a rank-2 tensor,
$T_{\mu\nu}$, the electromagnetic tensor, and gamma matrices, being addressed
in its axial and non-axial Hermitian versions, and also comprising general
possibilities. We then use the electron's anomalous magnetic dipole moment and
electron electric dipole moment measurements to reach upper bounds of $1$ part
in $10^{20}$ and $10^{25}$ (eV )$^{-1}$.

\end{abstract}

\maketitle

\section{Introduction}

The Standard Model (SM) structure allows for $C$, $P$ and $T$ violations (and
combinations) as long as the $CPT$ symmetry is kept unharmed. Concerning these
symmetries, it is fundamental to test their validity in any possible way.
Among the most important tests is the search for the electric dipole moment
(EDM). Its interaction term, in a nonrelativistic formulation, has the form
$d(\boldsymbol{\sigma}\cdot\mathbf{E}),$ in which $\mathbf{E}$ is the electric
field, $\boldsymbol{\sigma}$, the spin operator and $d$, the modulus of the
electric dipole moment (EDM). This interaction violates both $P$ and $T$
symmetries, $P(\boldsymbol{\sigma}\cdot\mathbf{E})\rightarrow
-(\boldsymbol{\sigma}\cdot\mathbf{E})$, $T(\boldsymbol{\sigma}\cdot
\mathbf{E})\rightarrow-(\boldsymbol{\sigma}\cdot\mathbf{E})$, but preserves
$C$, so the $CPT$ symmetry is not lost. The EDM magnitude $d$, according to
the SM \cite{SMEDM1,SMEDM2}, is $\approx10^{-38}\,e\cdot{\text{cm, }}$ while
the experimental measurements have been improved \cite{Measure1}%
,\cite{Measure2}, reaching the level $\approx10^{-31}\,e\cdot{\text{cm}}$ very
recently \cite{Baron}. The gap between the experimental landmark and the
theoretical prediction by a factor $10^{7}$ may look discouraging, but it also
means that any detection above the SM prediction could indicate New Physics,
that is, more sources of $CP$ violation. All this could play an interesting
role in explaining the matter-antimatter asymmetry, as the connection of
axions and the strong CP problem \cite{Axion}. By another route, EDM
experimental data can be used to set stringent bounds on theories that predict
this kind of effect.

The SM electrodynamics can be provided with EDM by introducing the term
$id(\bar{\psi}\sigma_{\mu\nu}\gamma_{5}F^{\mu\nu}\psi)$
\cite{EDM1,EDM2,LeptonEDM}, where $\psi$ represents a Dirac spinor. However,
there are other {mechanisms} for introducing EDMs on the SM framework. One of
these is to consider background fields, which interact with the spinor and
electromagnetic fields via nonminimal couplings. These background fields
induce preferred directions in spacetime, violating the Lorentz symmetry but
not necessarily harming $CPT$; a few scenarios involving breaking of Lorentz or
$CPT$ symmetries are discussed in Refs. \cite{Chaichian,Greenberg}. In Ref.
\cite{Pospelov}, the possible generation of EDM by several Lorentz-violating
(LV) dimension-5 interaction terms ($c^{\nu}\bar{\psi}\gamma^{\mu}F_{\mu\nu
}\psi,$ $d^{\nu}\bar{\psi}\gamma^{\mu}\gamma_{5}F_{\mu\nu}\psi,$ $f^{\nu}%
\bar{\psi}\gamma^{\mu}\tilde{F}_{\mu\nu}\psi,$ $g^{\nu}\bar{\psi}\gamma^{\mu
}\gamma_{5}\tilde{F}_{\mu\nu}\psi$) was pondered and used as a key factor for
constraining their respective magnitudes.

These $CPT$-odd terms were also studied in Ref. \cite{Stadnik}, regarding
their contribution to the anomalous magnetic moment (MDM). In this work, it
was performed an analysis involving the splitting of the $g$ factors of a
fermion and an antifermion, and bounds were set. As a matter of fact,
investigations concerning the muon's anomalous MDM have shown that its
experimental value already deviates from the QED prediction
\cite{Muon1,Muon2,Muon3}. Since muon-related experiments are about $400$ times
more sensitive to New Physics behavior (and probe mass scales around $20$
times higher), these searches motivate the proposal of alternative theories
\cite{LeptonEDM}, including the ones that allow for Lorentz and $CPT$
violations. In this context, there are investigations about LV effects on the
muon MDM \cite{Gomes} and on the neutron EDM \cite{Altarev}. $CPT$-even and LV
one-loop contributions to lepton EDM, induced by the SME fermion term,
$d_{\mu\nu}\bar{\psi}\gamma_{5}\gamma^{\mu}\psi,$\ were carried out in Ref.
\cite{Haghig}. CPT-odd LV effects on lepton EDM were addressed in Ref. \cite{NEW}.

The investigation of Lorentz symmetry violation is indeed a rich line of
research, much developed in the framework of the Standard Model extension
(SME) \cite{Colladay,fermion,CPT,fermion2,KM1,photons1}, whose developments
have scrutinized the Lorentz-violating effects in distinct physical systems
and served to state tight upper bounds on the LV coefficients, including
photon-fermion interactions \cite{Vertex}\textbf{\ }and electroweak processes
\cite{EW1},\cite{EW2}. Beyond the minimal SME, there is its nonminimal
extension encompassing nonminimal couplings with higher-order derivatives
\cite{NMSME}. Other models comprising higher-dimension operators \cite{Reyes}
and higher derivatives \cite{HD} have also been taken forward.

It is worth to mention that the $CPT$-odd term $g^{\nu}\bar{\psi}\gamma^{\mu
}\gamma_{5}\tilde{F}_{\mu\nu}\psi$ was first proposed in Ref. \cite{NM1} by
means of the nonminimal derivative, $D_{\mu}=\partial_{\mu}+ieA_{\mu}%
+i\frac{\lambda}{2}\epsilon_{\mu\lambda\alpha\beta}g^{\lambda}F^{\alpha\beta}%
$, defined in the context of the Dirac equation, $(i\gamma^{\mu}D_{\mu}%
-m)\Psi=0 $, where $g^{\mu}$ can be identified with the Carroll-Field-Jackiw
four-vector, $(k_{AF})^{\mu}=($v$_{0},\mathbf{v),}$ and $\lambda$ is the
coupling constant. This coupling has been studied in numerous aspects
\cite{NM3,NMmaluf,NMbakke}, including the radiative generation of $CPT$-odd LV
terms \cite{Radio}. See also Ref. \cite{NMABC} and the references therein.

We have recently investigated a dimension-five $CPT$-even nonminimal coupling
\cite{FredeNM1}, and its axial version \cite{Jonas1}, in the context of the
Dirac equation, implemented by means of the extended covariant derivatives,
\begin{align}
D_{\mu}  &  =\partial_{\mu}+ieA_{\mu}+\frac{\lambda}{2}(K_{F})_{\mu\nu
\alpha\beta}\gamma^{\nu}F^{\alpha\beta},\label{cov_even}\\
D_{\mu}  &  =\partial_{\mu}+ieA_{\mu}+i\frac{\lambda_{A}}{2}(K_{F})_{\mu
\nu\alpha\beta}\gamma_{5}\gamma^{\nu}F^{\alpha\beta}, \label{cov_even2}%
\end{align}
where $(K_{F})_{\mu\nu\alpha\beta}$\ is the $CPT$-even tensor of the SME
electrodynamics. They are associated with the CPT-even dimension-five
couplings deprived of higher derivatives,%
\begin{align}
&  \lambda\overline{\Psi}(K_{F})_{\mu\nu\alpha\beta}\sigma^{\mu\nu}%
F^{\alpha\beta}\Psi,\nonumber\label{NMcouplings}\\
& \\
&  \lambda_{A}\overline{\Psi}(K_{F})_{\mu\nu\alpha\beta}\gamma_{5}\sigma
^{\mu\nu}F^{\alpha\beta}\Psi,\nonumber
\end{align}
which are not contained in the broader nonminimal extension of the SME
developed in Refs. \cite{NMSME}.

In the Dirac equation, these couplings provide a nonrelativistic Hamiltonian
endowed with contributions to the EDM and to the MDM, which rendered upper
bounds on the LV parameters at the level of $1$ part in $10^{20}$ (eV)$^{-1}$
and $1$ part in $10^{24}$ (eV)$^{-1}$, respectively. Related studies arguing
the generation of topological phases have been reported as well \cite{Bakke2}.

In this letter, we propose some general dimension-five $CPT$-even nonminimal
couplings, composed of a rank-2 LV tensor, $T_{\mu\nu}$, in the context of the
Dirac equation, and also not contained in the nonminimal SME extension
proposed in Ref. \cite{NMSME}. Initially, the following extensions in the
covariant derivative are conceived,
\begin{align}
D_{\mu} &  =\text{$\partial$}_{\mu}+ieA_{\mu}+i\lambda_{1}T_{\mu\nu}%
F^{\nu\beta}\Gamma_{\beta},\\
D_{\mu} &  =\text{$\partial$}_{\mu}+ieA_{\mu}+i\lambda_{1}T_{\mu\nu}%
F_{\alpha\beta}\gamma^{\nu}\Gamma^{\alpha\beta},
\end{align}
where $\Gamma_{\beta}=\gamma_{\beta},\gamma_{\beta}\gamma_{5}$ stand for the
axial and non axial forms, and $\Gamma^{\alpha\beta}=\sigma^{\alpha\beta}.$
These terms engender contributions to the Dirac Lagrangian, which are
reorganized in such a way to be Hermitian. Table 1 contains four types of
possible extensions, followed by their respective Hermitian versions.

\begin{table}[h]
\centering%
\begin{tabular}
[c]{|c|c|c|c|}\hline
Coupling & Hermitian & $EDM$ & $MDM$\\\hline
$\lambda_{1}\bar{\Psi}T_{\mu\nu}F^{\nu\beta}{\gamma^{\mu}}\gamma_{\beta}%
\gamma^{5}\Psi$ & $no$ & $-$ & $-$\\\hline
$i\lambda_{1}\bar{\Psi}T_{\mu\nu}F^{\nu}{}_{\beta}\sigma^{\mu\beta}\gamma
^{5}\Psi$ & $yes$ & $yes$ & $"yes"$\\\hline
$\lambda_{1}^{\prime}\bar{\Psi}T_{\mu\nu}F_{\text{ }\beta}^{\nu}{\gamma^{\mu}%
}\gamma^{\beta}\Psi$ & $no$ & $-$ & $-$\\\hline
$i\lambda_{1}^{\prime}\bar{\Psi}T_{\mu\nu}F^{\nu}{}_{\beta}\sigma^{\mu\beta
}\Psi$ & $yes$ & $"yes"$ & $yes$\\\hline
$\lambda_{3}\bar{\Psi}T_{\alpha\nu}F_{\mu\beta}{\gamma^{\mu}}\gamma^{\beta
}\gamma^{\alpha}\gamma^{\nu}\Psi$ & $no$ & $-$ & $\ -$\\\hline
\multicolumn{1}{|l|}{$\lambda_{3}\bar{\Psi}\left(  T_{\alpha\nu}F_{\mu\beta
}+T_{\mu\beta}F_{\alpha\nu}\right)  \sigma^{\mu\beta}\sigma^{\alpha\nu}\Psi$}
& \multicolumn{1}{|l|}{$\ \ \ \ \ yes$} & \multicolumn{1}{|l|}{$\ \ no$} &
$no$\\\hline
$\lambda_{4}\bar{\Psi}\left(  T_{\alpha\nu}F_{\mu\beta}-T_{\mu\beta}%
F_{\alpha\nu}\right)  \sigma^{\mu\beta}\sigma^{\alpha\nu}\Psi$ & $yes$ &
$"yes"$ & $"yes"$\\\hline
\end{tabular}
\caption{EDM and MDM contributions arisen from possible LV extensions and
their respective Hermitian versions. Here, $"yes"$ means the
conditional possibility of measuring EDM or MDM in a non conventional set
up.}%
\end{table}

After presenting each case and accessing the nonrelativistic regime of the
resulting Hermitian Dirac equation, we use the electron's MDM and EDM data to
impose limits on the magnitude of the nonminimal LV terms at the level of
until $1$ part in $10^{25}$ (eV)$^{-1}$ or $1$\ part in $10^{16}$%
\ (GeV)$^{-1}.$

\section{An axial and Non axial CPT-even LV nonminimal coupling in the Dirac
equation}

In the context of the Dirac equation, we examine general CPT-even nonminimal
couplings not composed of higher derivative orders, representing scenarios
where an unusual electromagnetic interaction between fermions and photons is
induced by a fixed rank-2 tensor, without any symmetry (in principle).

\subsection{The axial nonminimal coupling}

The starting point is the proposal of a general $CPT$-even nonminimal term in
the form of an extension in the covariant derivative,
\begin{equation}
D_{\mu}=\text{$\partial$}_{\mu}+ieA_{\mu}+i\lambda_{1}T_{\mu\nu}F^{\nu\beta
}\gamma^{5}\gamma_{\beta}, \label{covader}%
\end{equation}
with $T_{\mu\nu}$ being a rank-2 arbitrary, dimensionless, LV tensor, in
principle without any symmetry. {It generates a dimension-five operator,
coupling the fields in this modified quantum electrodynamics, i.e.,
$\lambda_{1}\overline{\Psi}T_{\mu\nu}F^{\nu\beta}\gamma^{\mu}\gamma^{5}%
\gamma_{\beta}\Psi$. The coupling constant $\lambda_{1}$ is a real parameter
with mass-dimension equals to $-1$. In order to yield a Hermitian
contribution}, one should consider the new nonminimal covariant derivative as
\begin{equation}
D_{\mu}=\text{$\partial$}_{\mu}+ieA_{\mu}-\frac{i}{2}\lambda_{1}\left(
T_{\mu\nu}F^{\nu\beta}-T_{\beta\nu}F_{\text{ }\mu}^{\nu}\right)  \gamma
_{\beta}\gamma^{5}. \label{covader1}%
\end{equation}
Accordingly, the modified Dirac equation reads,
\begin{equation}
\left[  i\gamma^{\mu}\partial_{\mu}-e\gamma^{\mu}A_{\mu}-i\lambda_{1}T_{\mu
\nu}F_{\text{ \ }\beta}^{\nu}\sigma^{\mu\beta}\gamma^{5}-m\right]  \Psi=0,
\label{DiracM1}%
\end{equation}
with $i\lambda_{1}\overline{\Psi}T_{\mu\nu}F_{\text{ \ }\beta}^{\nu}%
\sigma^{\mu\beta}\gamma^{5}\Psi$ representing the corresponding LV interaction
between spinors and the electromagnetic field.

Here, $F_{0j}=E^{j},F_{mn}=\epsilon_{mnp}B_{p}$, while $\sigma^{0j}%
=i\alpha^{j},$ $\sigma^{ij}=\epsilon_{ijk}\Sigma^{k},$ with the Dirac matrices
given as
\[
\alpha^{i}=\left(
\begin{array}
[c]{cc}%
0 & \sigma^{i}\\
\sigma^{i} & 0
\end{array}
\right)  ,\text{ \ \ \ \ }\Sigma^{k}=\left(
\begin{array}
[c]{cc}%
\sigma^{k} & 0\\
0 & \sigma^{k}%
\end{array}
\right)  ,\text{ \ \ \ }\gamma^{5}=\left(
\begin{array}
[c]{cc}%
0 & 1\\
1 & 0
\end{array}
\right)  ,
\]
in which $\sigma^{i}$ stands for the Pauli matrices. In the momentum
coordinates, $i\partial_{\mu}\rightarrow p_{\mu},$ the corresponding Dirac
equation can be written as
\begin{equation}
i\partial_{t}\Psi=\left[  \boldsymbol{\alpha}\cdot\boldsymbol{\pi}%
+eA_{0}+m\gamma^{0}+H_{\text{LV}}\right]  \Psi, \label{DiracM1C}%
\end{equation}
where $\boldsymbol{\pi}=\left(  \mathbf{p-eA}\right)  $ is the canonical
momentum and
\begin{equation}
H_{\text{LV}}=i\lambda_{1}T_{\mu\nu}F_{\text{ \ }\beta}^{\nu}\gamma^{0}%
\sigma^{\mu\beta}\gamma^{5},
\end{equation}
is the LV piece,\textsl{\ }which, inserted in the Dirac equation
(\ref{DiracM1C}), implies%
\begin{align}
i\partial_{t}\Psi &  =\left[  \boldsymbol{\alpha}\cdot\boldsymbol{\pi}%
+eA_{0}+m\gamma^{0}\frac{{}}{{}}\right. \nonumber\\
&  -\lambda_{1}T_{00}E^{j}\gamma^{0}\Sigma^{j}+\lambda_{1}T_{ij}E^{i}%
\gamma^{0}\Sigma^{j}\nonumber\\
&  -\lambda_{1}\epsilon_{ijk}T_{0i}B^{k}\gamma^{0}\Sigma^{j}+i\lambda
_{1}\epsilon_{jki}T_{j0}E^{k}\gamma^{i}\nonumber\\
&  \left.  \frac{{}}{{}}+i\lambda_{1}TB^{k}\gamma^{k}%
-i\lambda_{1}T_{jk}B^{j}\gamma^{k})\right]  \Psi,
\end{align}
where $T=\mbox{Tr}(T_{ij})$. The presence of the matrix,
$\gamma^{0}$, in the terms $T_{00}\gamma^{0}E^{i}\Sigma^{i},$ $T_{ij}%
\gamma^{0}E^{j}\Sigma^{i},$ indicates effective induction of EDM by
electromagnetic interaction, once it will contribute to the energy of the
system, circumventing the Schiff's theorem \textbf{(}see Refs.
\cite{LeptonEDM},\cite{Jonas1}\textbf{)}. As a consequence, one can use the
electron EDM data to constrain the magnitude of the coefficients
$T_{00},T_{ij}.$

In order to investigate the role played by this nonminimal coupling, we should
evaluate the nonrelativistic limit of the Dirac equation. We begin by writing
the LV Hamiltonian of Eq. (\ref{DiracM1C}) in a matrix form,%
\begin{equation}
H_{\text{LV}}=%
\begin{pmatrix}
H_{11} & -H_{12}\\
H_{12} & -H_{11}%
\end{pmatrix}
, \label{Hmatrix}%
\end{equation}
with%
\begin{align}
H_{11}  &  =\lambda_{1}(-T_{00}\left(  \boldsymbol{\sigma}\cdot\boldsymbol{E}%
\right)  -T_{0i}(\boldsymbol{\sigma}\times\boldsymbol{B})^{i}+T_{ij}%
E^{j}\sigma^{i}),\label{H11}\\
H_{12}  &  =i\lambda_{1}(-\epsilon_{ijk}T_{i0}\sigma^{k}E^{j}-T_{ii}\left(
\boldsymbol{\sigma}\cdot\boldsymbol{B}\right)  +T_{ij}\sigma^{j}B^{i}).
\label{H12}%
\end{align}
Labeling the small $\left(  \chi\right)  $\ and large $\left(  \psi\right)
$\ components of the spinor$\ \Psi$, the Dirac equation (\ref{DiracM1C})
provides two $2$-component equations,%
\begin{align}
\lbrack E-eA_{0}-m-H_{11}]\psi-[\boldsymbol{\sigma}\cdot\boldsymbol{\pi
}-H_{12}]\chi &  =0,\\
\lbrack\boldsymbol{\sigma}\cdot\boldsymbol{\pi}+H_{12}]\psi-[E-eA_{0}%
+m+H_{11}]\chi &  =0.
\end{align}

After some algebraic evaluations, at first order in the Lorentz violating
parameters, the nonrelativistic Hamiltonian (for uniform fields) is
\begin{align}
H_{\text{NR}}  &  =\frac{1}{2m}\left[  \boldsymbol{\pi}{}^{2}%
-e\boldsymbol{\sigma}\cdot\boldsymbol{B}\right]  +eA_{0}\nonumber\\
&  -\lambda_{1}\left(  T_{00}\boldsymbol{\sigma}\cdot\boldsymbol{E}%
+T_{0i}(\boldsymbol{\sigma}\times\boldsymbol{B})^{i}-T_{ij}E^{j}\sigma
^{i}\right) \nonumber\\
&  +\frac{1}{2m}\left[  (\boldsymbol{\sigma}\cdot\boldsymbol{\pi)}%
H_{12}-H_{12}(\boldsymbol{\sigma}\cdot\boldsymbol{\pi})\right]  . \label{HNR3}%
\end{align}

Examining this Hamiltonian, we notice that it is able to generate electric
dipole moment for the electron at tree level, by the terms $\lambda_{1}%
T_{00}\left(  \boldsymbol{\sigma}\cdot\boldsymbol{E}\right)  ,$ $\lambda
_{1}T_{ij}E^{j}\sigma^{i},$ but no anomalous magnetic moment. The EDM terms
could also be seen as generator of a kind of electric Zeeman effect.

It is interesting to mention that only the symmetric part of the tensor
$T_{\mu\nu}$ is able to induce EDM, once the antisymmetric part has null main
diagonal, avoiding the appearance of a term containing $\left(
\boldsymbol{\sigma}\cdot\boldsymbol{E}\right)  $ in $H_{\text{NR}}.$

We also discuss the behavior of the modified Dirac equation (\ref{DiracM1})
and (\ref{DiracM1C}) under the discrete symmetries $C,P,T$. \ Table I displays
the response of the elements $\lambda T_{00},\lambda T_{0i},\lambda
T_{ij},\lambda T_{ii}$ under the $C,P,T$ operators. We can also observe that
$\lambda T_{00},$ $\lambda T_{ij}$ are $P$-odd and $T$-odd, compatible with
the EDM character.

The Hamiltonian (\ref{HNR3}) contains LV EDM terms, $\lambda T_{00}\left(
\boldsymbol{\sigma}\cdot\boldsymbol{E}\right)  ,\lambda T_{ij}E^{j}\sigma
^{i},$ on which one can set stringent upper bounds using experimental data. We
begin with
\begin{equation}
\lambda_{1}T_{00}\left(  \boldsymbol{\sigma}\cdot\boldsymbol{E}\right)  ,
\end{equation}
where we notice that $\lambda T_{00}$ plays the role of the electron EDM
magnitude, $\left\vert \mathbf{d}_{e}\right\vert .$ The magnitude of
$\mathbf{d}_{e}$ has been constrained with increasing precision
\cite{EDM1,Measure1,Measure2}, recently reaching the level $\left\vert
\mathbf{d}_{e}\right\vert \leq8.7\times10^{-31}$ e.m or $\left\vert
\mathbf{d}_{e}\right\vert \leq3.8\times10^{-25}(\mbox{eV})^{-1}$ \cite{Baron}.
Considering this experimental measure, we attain the following upper bound:
\begin{equation}
\left\vert \lambda_{1}T_{00}\right\vert \leq3.8\times10^{-16}(\mbox{GeV})^{-1}%
. \label{bound2}%
\end{equation}

As for the term, $\lambda_{1}T_{ij}E^{j}\sigma^{i},$ it can be written as
$\lambda(T/3)\left(  \boldsymbol{\sigma}\cdot\mathbf{E}\right)  ,$ where
{$T=\mbox{Tr}(T_{ij})$}. Similarly, we state on the trace:%
\begin{equation}
\left\vert \lambda_{1}T\right\vert \leq1.1\times10^{-15}(\mbox{GeV})^{-1},
\label{bound2b}%
\end{equation}
or for a specific component $\lambda_{1}T_{ii}$,
\begin{equation}
\left\vert \lambda_{1}T_{ii}\right\vert \leq3.8\times10^{-16}(\mbox{GeV})^{-1}%
. \label{bound2c}%
\end{equation}
These upper limits are among the best ones to be stated over dimension five
CPT-even LV nonminimal coefficients (not involving higher derivatives).
Considering a situation in which there is impossibility of identifying the LV
effects stemming from $T_{00}$\ and $T_{ii},$\ the EDM interaction is taken
as
\begin{equation}
\lambda_{1}\left(  T_{00}\boldsymbol{\sigma}\cdot\boldsymbol{E}-T_{ij}%
E^{j}\sigma^{i}\right)  =\lambda_{1}\left(  T_{00}-T_{ii}\right)  \left(
\boldsymbol{\sigma}\cdot\boldsymbol{E}\right)  ,
\end{equation}
reflecting the case when the diagonal elements $T_{ii}$\ are equal. It yields
the following upper bound:
\begin{equation}
\lambda_{1}\left\vert T_{00}-T_{ii}\right\vert \leq3.8\times10^{-16}%
(\mbox{GeV})^{-1}. \label{bound2dd}%
\end{equation}
For a specific value of $i$, it holds $\left(  \text{tr}T_{\mu\nu}\right)
=(T_{00}-T)=(T_{00}-3T_{ii})$\ and $\left(  \text{tr}T_{\mu\nu}\right)
=(T_{00}-T_{ii})-2T_{ii}.$\ Considering that the blocks $2T_{ii}$\ and
$(T_{00}-T_{ii})$\ are limited as stated in Eqs. (\ref{bound2c},
\ref{bound2dd}), the trace element can be restrained as
\begin{equation}
\left\vert \lambda_{1}\left(  \text{tr}T_{\mu\nu}\right)  \right\vert
\leq7.6\times10^{-16}(\mbox{GeV})^{-1}. \label{Ttrace}%
\end{equation}

We still comment on a nonconventional interpretation that allows to constrain
the component $\lambda_{1}T_{0i}(\sigma\times B)^{i}=\lambda_{1}%
(\epsilon_{ijk}T_{0i}\sigma^{j})B^{k}$\ of Hamiltonian (\ref{HNR3}), which can
be read as $\lambda_{1}\boldsymbol{\tilde{\sigma}}\cdot
\mathbf{B}=\lambda_{1}\tilde{\sigma}^{k}B^{k}$, where $\tilde{\sigma
}^{k}=\epsilon_{ijk}T_{0i}\sigma^{j}$\ is a kind of "rotated" spin operator
yielding a "rotated" magnetic moment generated by the Lorentz violating
background.This term can only be constrained with MDM data if one
conceives a non conventional experimental set up able to measure a non null
spin expectation value, $\left\langle S_{i}\right\rangle ,$ in a direction
orthogonal to the applied magnetic field ($B$). In this specific context, the
same procedure developed in Eqs. (\ref{MDMway1}) and (\ref{MDMway2}) could
imply the upper bound
\begin{equation}
\left\vert \lambda_{1}T_{0i}\right\vert \leq5.5\times10^{-11}%
\,(\mbox{GeV})^{-1}.\label{MDMT0i}%
\end{equation}

Moreover, the bounds (\ref{bound2c}) and (\ref{MDMT0i}) are subject to
sidereal variations, since the LV background field is approximately constant
only on the Sun's reference frame (RF). It is necessary, therefore, to bring
these bounds to the Earth-located Lab's RF, at the colatitude $\chi$, rotating
around the Earth's axis with angular velocity $\Omega$. For experiments up to
a few weeks long, the transformation law for a rank-2 tensor, $A_{\mu\nu}$,
according to Refs. \cite{Jonas1,Sideral} is merely a spatial rotation,\textbf{
}%
\begin{equation}
A_{\mu\nu}^{\text{(Lab)}}=\mathcal{R}_{\mu\alpha}\mathcal{R}_{\nu\beta
}A_{\alpha\beta}^{\text{(Sun)}},\label{rotationLaw}%
\end{equation}
where
\begin{equation}
R_{\mu\nu}=%
\begin{pmatrix}
1 & 0 & 0 & 0\\
0 & \cos\chi\cos\Omega t & \cos\chi\sin\Omega t & -\sin\chi\\
0 & -\sin\Omega t & \cos\Omega t & 0\\
0 & \sin\chi\cos\Omega t & \sin\chi\sin\Omega t & \cos\chi\label{Rotation}%
\end{pmatrix}
,
\end{equation}
in which the first line and column were included. The $T_{00}$ (as its time
average) does not change, just as the spatial trace, $T^{\text{(Lab)}}=\text{Tr}(T_{ij})^{\text{(Lab)}}%
$. However, any specific main diagonal component $T_{ii}$ (for a particular
$i$ value) in (\ref{bound2c}) is modified, having time average given as
\begin{align}
\langle T_{ii}^{\text{(Lab)}}\rangle &  =\langle(\mathcal{R}_{i1})^{2}\rangle
T_{11}^{\text{(Sun)}}+\langle(\mathcal{R}_{i2})^{2}\rangle T_{22}%
^{\text{(Sun)}}\nonumber\\
&  +\langle(\mathcal{R}_{i3})^{2}\rangle T_{33}^{\text{(Sun)}},
\end{align}
because $\langle R_{i1}R_{i2}\rangle=\langle R_{i1}R_{i3}\rangle=\langle
R_{i2}R_{i3}\rangle=0$, for any $i$, due to their dependence on $\sin\Omega
t$\ and $\cos\Omega t$. \ As for the bound (\ref{MDMT0i}), it transforms as a
regular vector, and its time average becomes\textbf{ }%
\begin{equation}
\langle T_{0i}^{\text{(Lab)}}\rangle=(-\delta_{i1}\sin\chi+\delta_{i3}\cos
\chi)T_{03}^{\text{(Sun)}},
\end{equation}
since $\langle R_{i1}\rangle=\langle R_{i2}\rangle=0$. Hence, the bounds
(\ref{bound2b}), (\ref{bound2c}) and (\ref{MDMT0i}) on the Lab's RF are,
respectively,
\begin{equation}
|\lambda_{1}\langle T^{\text{(Sun)}}\rangle|\leq1.1\times10^{-15}%
(\mbox{GeV})^{-1},
\end{equation}%
\begin{align}
&  |\lambda_{1}\langle(\mathcal{R}_{i1})^{2}\rangle T_{11}^{\text{(Sun)}%
}+\lambda_{1}\langle(\mathcal{R}_{i2})^{2}\rangle T_{22}^{\text{(Sun)}%
}\nonumber\\
&  +\lambda_{1}\langle(\mathcal{R}_{i3})^{2}\rangle T_{33}^{\text{(Sun)}}%
|\leq3.8\times10^{-16}(\mbox{GeV})^{-1},
\end{align}%
\begin{equation}
\left\vert \lambda_{1}(-\delta_{i1}\sin\chi+\delta_{i3}\cos\chi)T_{03}%
^{\text{(Sun)}}\right\vert \leq5.5\times10^{-11}\,(\mbox{GeV})^{-1}.
\end{equation}

\subsection{Non axial nonminimal coupling}

Other possibility of coupling the fermion and electromagnetic fields by means
of a CPT-even nonminimal coupling involving a rank-2 tensor to be mentioned is
the Hermitian and non axial version of the coupling (\ref{covader1}), that
is,
\begin{equation}
D_{\mu}=\partial_{\mu}+ieA_{\mu}+\frac{\lambda_{1}^{\prime}}{2}\left(
T_{\mu\nu}F_{\text{ }\beta}^{\nu}-T_{\beta\nu}F_{\text{ \ }\mu}^{\nu}\right)
\gamma^{\beta}, \label{covader2}%
\end{equation}
leading to the following Dirac equation,
\begin{equation}
\left[  i\gamma^{\mu}\partial_{\mu}-e\gamma^{\mu}A_{\mu}+\lambda_{1}T_{\mu\nu
}F_{\text{ \ }\beta}^{\nu}\sigma^{\mu\beta}-m\right]  \Psi=0. \label{DiracM2}%
\end{equation}
The corresponding LV Hamiltonian, $H_{\text{LV}}^{\prime}=-\lambda_{1}%
^{\prime}T_{\mu\nu}F_{\text{ \ }\beta}^{\nu}\gamma^{0}\sigma^{\mu\beta},$ can
be expressed as%
\begin{align}
H_{\text{LV}}^{\prime}  &  =-i\lambda_{1}^{\prime}T_{00}E^{i}\gamma
^{i}-i\lambda_{1}^{\prime}T_{0i}\epsilon_{ijb}B^{b}\gamma^{j}-\lambda
_{1}^{\prime}T_{i0}\epsilon_{ija}E^{j}\gamma^{0}\Sigma^{a}\nonumber\\
&  +i\lambda_{1}^{\prime}T_{ij}E^{j}\gamma^{i}-\lambda_{1}^{\prime}T_{ii}%
B^{a}\gamma^{0}\Sigma^{a}+\lambda_{1}^{\prime}T_{ia}B^{i}\gamma^{0}\Sigma^{a},
\end{align}
or in the matrix form (\ref{Hmatrix}), with components:
\begin{align}
H_{11}^{\prime}  &  =-\lambda_{1}^{\prime}(T_{i0}E^{j}\epsilon_{ijk}\sigma
^{k}+T_{ii}\sigma^{p}B^{p}-T_{ij}B^{i}\sigma^{j}),\\
H_{12}^{\prime}  &  =i\lambda_{1}^{\prime}(T_{00}E^{i}\sigma^{i}-T_{ij}%
E^{j}\sigma^{i}+T_{0i}\epsilon_{ijk}\sigma^{j}B^{k}).
\end{align}
The nonrelativistic Hamiltonian is%
\begin{align}
H_{\text{NR}}  &  =\frac{1}{2m}\left[  (\boldsymbol{p}-e\boldsymbol{A}){}%
^{2}-e\boldsymbol{\sigma}\cdot\boldsymbol{B}\right]  +eA_{0}\nonumber\\
&  +\lambda_{1}^{\prime}T_{i0}\left(  \boldsymbol{\sigma}\times\boldsymbol{E}%
\right)  ^{i}-\lambda_{1}^{\prime}T_{ii}\left(  \boldsymbol{\sigma}%
\cdot\boldsymbol{B}\right)  +\lambda_{1}^{\prime}T_{ij}\sigma^{j}%
B^{i}\nonumber\\
&  +\frac{1}{2m}[\left(  \boldsymbol{\sigma}\cdot\boldsymbol{\pi}\right)
H_{12}^{\prime}-H_{12}^{\prime}\left(  \boldsymbol{\sigma}\cdot\boldsymbol{\pi
}\right)  ]. \label{HNR5}%
\end{align}
\qquad

Clearly, the terms $T_{ii}\left(  \boldsymbol{\sigma}\cdot\boldsymbol{B}%
\right)  $\ and $T_{ij}\sigma^{j}B^{i}$ are MDM contributions, associated only
with the symmetric part of the tensor $T_{\mu\nu}$, whose magnitude can be
limited by the known experimental error in the MDM.

The electron's magnetic moment is $\boldsymbol{\mu}=-g\mu_{B}\boldsymbol{S},$
with $\mu_{B}=e/2m$ being the Bohr magneton and $g=2(1+a)$ being the
gyromagnetic factor, with $a=\alpha/2\pi\simeq0.00116$ representing the
deviation from the usual case, $g=2$. The magnetic interaction is $H^{\prime
}=-\mu_{B}g\left(  \boldsymbol{S}\cdot\mathbf{B}\right)  $. The most precise
calculation up to date of $a$ is found in Ref. \cite{anomalousMP}.
Experimentally, precise measurements \cite{Gabrielse} reveal that the error on
the electron MDM is at the level of $2.8$\ parts in $10^{13},$\ that is,
$\Delta a\leq2.8\times10^{-13}.$ Hamiltonian (\ref{HNR5}) possesses tree-level
LV\ contributions to the usual $g=2$\ gyromagnetic factor, that is%
\begin{equation}
\lambda_{1}^{\prime}T_{ii}\left(  \boldsymbol{\sigma}\cdot\boldsymbol{B}%
\right)  -\lambda_{1}^{\prime}\frac{T_{ii}}{3}\left(  \boldsymbol{\sigma}%
\cdot\boldsymbol{B}\right)  =\frac{2}{3}\lambda_{1}^{\prime}T_{ii}\left(
\boldsymbol{\sigma}\cdot\boldsymbol{B}\right)  ,\label{MDMway1}%
\end{equation}
which can not be larger than $\Delta a.$ The total magnetic interaction in Eq.
(\ref{HNR5}) is $\mu_{B}\left(  \boldsymbol{\sigma\cdot B}\right)  +\frac
{2}{3}\lambda_{1}^{\prime}T_{ii}\left(  \boldsymbol{\sigma}\cdot
\boldsymbol{B}\right)  $, so that this interaction assumes the form
\begin{equation}
\mu_{B}\left[  1+\frac{2}{3}\frac{2m}{e}\lambda_{1}^{\prime}T_{ii}\right]
\left(  \boldsymbol{\sigma}\cdot\boldsymbol{B}\right)  ,\label{MDMway2}%
\end{equation}
where $\frac{4m}{3e}\lambda_{1}^{\prime}T_{ii}$ stands for the tree-level LV
correction that should be smaller than $\Delta a$. This leads to the following
upper bound for the trace:
\begin{equation}
\left\vert \lambda_{1}^{\prime}T_{ii}\right\vert \leq3.5\times10^{-11}%
\,(\mbox{GeV})^{-1},\label{MDb1}%
\end{equation}
being less restrictive by a factor $\simeq10^{5}$ than the previous ones
derived by EDM.

On the Sun's RF, this bound is equally written as
\begin{equation}
\left\vert \lambda_{1}^{\prime}T_{ii}^{\text{(Sun)}}\right\vert \leq
3.5\times10^{-11}(\mbox{GeV})^{-1},
\end{equation}
once the trace element is invariant under the rotation (\ref{Rotation}). 

The behavior of the term $\lambda_{1}^{\prime}T_{i0}$\ of Hamiltonian
(\ref{HNR5}), under the discrete operations $C,P$\ and $T$\ (see Table II), is
compatible with the EDM properties. Accordingly, the term $\lambda_{1}%
^{\prime}T_{i0}\left(  \boldsymbol{\sigma}\times\boldsymbol{E}\right)
^{i}=\lambda_{1}^{\prime}T_{i0}\epsilon_{ijk}\sigma^{j}E^{k}$ could be
considered as EDM if we take
\begin{equation}
\lambda_{1}^{\prime}T_{i0}\left(  \boldsymbol{\sigma}\times\boldsymbol{E}%
\right)  ^{i}=\lambda_{1}^{\prime}\tilde{\sigma}^{k}E^{k},\label{Erotated1}%
\end{equation}
\ with $\tilde{\sigma}^{k}=\epsilon_{ijk}T_{0i}\sigma^{j}$ implying a kind of
"rotated" EDM. Analogously to the magnetic case, it can only be limited with
EDM data if there is a non-conventional experimental device able to measure a
non-null spin expectation value, $\left\langle S_{i}\right\rangle ,$ in a
direction orthogonal to the applied electric field ($E$). In this particular
situation, the procedure of Eq. (\ref{bound2}) could engender the upper bound
\begin{equation}
\left\vert \lambda_{1}^{\prime}T_{i0}\right\vert \leq3.8\times10^{-16}%
(\mbox{GeV})^{-1},\label{bErotated1}%
\end{equation}
which is also subject to sidereal variations, that is
\begin{equation}
\left\vert \lambda_{1}^{\prime}(-\delta_{i1}\sin\chi+\delta_{i3}\cos
\chi)T_{30}^{\text{(Sun)}}\right\vert \leq3.8\times10^{-16}\,(\mbox{GeV})^{-1}%
.
\end{equation}

\subsection{Comments and correspondences}

The classification of the LV coefficients of the axial and non-axial
couplings, including some other of the couplings to be examined in the next
section, under $C,P$ and $T$, is enclosed in Table II. \begin{table}[h]
\centering%
\begin{tabular}
[c]{|l|l|l|l|l|l|l|l|l|}\hline
& $\lambda_{1}T_{00}$ & $\lambda_{1}T_{0i}$ & $\lambda_{1}T_{ij}$ &
$\lambda_{1}^{\prime}T_{00}$ & $\lambda_{1}^{\prime}T_{0i}$ & $\lambda
_{1}^{\prime}T_{ij}$ & $\lambda_{4}T_{0i}$ & $\lambda_{4}T_{ij}$\\\hline
$C$ & $\ \ +$ & $\ \ +$ & $\ \ +$ & $\ \ +$ & $\ \ +$ & $\ \ +$ & $\ \ +$ &
$\ \ +$\\\hline
$P$ & $\ \ -$ & $\ \ +$ & $\ \ -$ & $\ \ +$ & $\ \ -$ & $\ \ +$ & \ \ $-$ &
$\ \ +$\\\hline
$T$ & $\ \ -$ & $\ \ +$ & $\ \ -$ & $\ \ +$ & $\ \ -$ & $\ \ +$ & $\ \ -$ &
$\ \ +$\\\hline
$CPT$ & $\ \ +$ & $\ \ +$ & $\ \ +$ & $\ \ +$ & $\ \ +$ & $\ \ +$ & $\ \ +$ &
$\ \ +$\\\hline
\end{tabular}
\caption{Classification under $C,P,T$ for the coefficients of the axial and
non-axial CPT-even {nonminimal} couplings. }%
\end{table}

We should comment on a partial equivalence: in the case the tensor $T_{\mu\nu
}$ is symmetric and traceless, the nonminimal couplings (\ref{covader1}) and
(\ref{covader2}) recover part of the nonminimal coupling first analyzed in
Refs. \cite{Jonas1} and \cite{FredeNM1}, respectively. Indeed, we begin by
taking a symmetric and traceless tensor, $\kappa_{\nu\beta}$ ($\kappa_{\ \nu
}^{\nu}=0\rightarrow\kappa_{00}=\kappa_{jj}$), linked to the $(K_{F})$ tensor
by $\kappa_{\nu\beta}$\ $=(K_{F})_{\text{ \ }\nu\alpha\beta}^{\alpha}%
,$\ and\textbf{ }%
\begin{equation}
(K_{F})_{\mu\nu\alpha\beta}=\frac{1}{2}\left(  g_{\mu\alpha}\kappa_{\nu\beta
}-g_{\mu\beta}\kappa_{\nu\alpha}+g_{\nu\beta}\kappa_{\mu\alpha}-g_{\nu\alpha
}\kappa_{\mu\beta}\right)  . \label{presc1}%
\end{equation}
Replacing\ this latter expression in the couplings $\overline{\Psi}\lambda
_{A}(K_{F})_{\mu\nu\alpha\beta}\sigma^{\mu\nu}\gamma^{5}F^{\alpha\beta}\Psi
$\ and $\overline{\Psi}\lambda(K_{F})_{\mu\nu\alpha\beta}\sigma^{\mu\nu
}F^{\alpha\beta}\Psi,$\ yields $\overline{\Psi}\lambda\kappa_{\mu\beta
}F_{\text{ }\nu}^{\beta}\sigma^{\mu\nu}\gamma^{5}\Psi$\ and $\overline{\Psi
}\lambda\kappa_{\mu\beta}F_{\text{ }\nu}^{\beta}\sigma^{\mu\nu}\Psi
,$\ respectively, which have the same form of the operators appearing in the
Dirac equations (\ref{DiracM1}) and (\ref{DiracM2}). This equivalence holds
only if the tensor $T_{\mu\nu}$\ is symmetric and traceless,
however. We point out that the present development is broader, since the
tensor $T_{\mu\nu}$ has not a definite symmetry (in principle). We also
comment on the possible correspondence between the previous upper bounds on
the couplings $\lambda_{A}(K_{F})$\ and $\lambda(K_{F})$,\ achieved in Refs.
\cite{Jonas1} and \cite{FredeNM1}, and the present ones, as shown in Table III.

\begin{table}[h]
\centering%
\begin{tabular}
[c]{|l|l|l|}\hline
& $\ \ \ \ \ \ \ \ \ \ \ \ \ \ \ \ \ \ \lambda(K_{F})$ &
\ \ \ \ \ \ \ \ Counterpart\\\hline
$MDM$ & $\ \ \lambda\left(  \kappa_{HB}\right)  _{33}\leq2.3\times10^{-20}\,$
& \ $\left\vert \lambda_{1}^{\prime}T_{ii}\right\vert \leq3.5\times10^{-20}%
$\\\hline
$EDM$ & $\lambda\left(  \kappa_{HE}\right)  _{11}\leq3.8\times10^{-25}$ &
\ \ \ \ \ \ \ \ \ \ \ \ \ \ \ no\\\hline
& $\ \ \ \ \ \ \ \ \ \ \ \ \ \ \ \ \lambda_{A}(K_{F})$ & \\\hline
$MDM$ & $\left\vert \lambda_{A}\left(  \kappa_{DB}\right)  _{33}\right\vert
\leq2.3\times10^{-20}\,$ & \ \ \ \ \ \ \ \ \ \ \ \ \ \ \ no\\\hline
$EDM$ & $\left\vert \lambda_{A}(\kappa_{DE})_{ii}\right\vert \leq
1.1\times10^{-24}$ & $\left\vert \lambda_{1}\left(  \text{tr}T_{\mu\nu
}\right)  \right\vert \leq3.8\times10^{-25}$\\\hline
\end{tabular}
\caption{Previous upper bounds for the axial and non-axial CPT-even
{nonminimal} couplings. All bounds expressed in terms of $(\mbox{eV})^{-1}.$}%
\end{table}In Table III, we notice that the first and fourth bound find a
correspondence with the present nonminimal couplings. The first one involves
the component $T_{00},$\ which can be justified by the relation%
\begin{equation}
\left(  \kappa_{DE}\right)  ^{jk}=\delta^{jk}\kappa_{00}-\kappa^{jk}\mathbf{,}%
\end{equation}
with $\left(  \kappa_{DE}\right)  ^{jj}=$\ $2\kappa_{00}/3$, for a specific
value of $j,$\ and tr$\left(  \kappa_{DE}\right)  =2\kappa_{00}.$ Concerning
the first bound of Table III, the relation%

\begin{equation}
\left(  \kappa_{HB}\right)  ^{jk}=-\delta^{jk}\kappa^{ll}+\kappa^{kj},
\end{equation}
leads to $\left(  \kappa_{HB}\right)  ^{jj}=-2\kappa_{00}/3$\ for a specific
value of $j,$\ and tr$\left(  \kappa_{HB}\right)  =-2\kappa_{ii}.$\ In this
sense, the first bound on $\ \lambda\left(  \kappa_{HB}\right)  _{33}$ in
Table III can be related to the bound (\ref{MDb1}) on $\lambda_{1}^{\prime
}T_{ii}.$

A possible correspondence to the terms involving $\left(  \kappa_{DB}\right)
,\left(  \kappa_{HE}\right)  $ in the models of Ref. \cite{Jonas1} takes place
by the appearance of the pieces $\lambda_{1}T_{0i}(\sigma\times B)^{i}$ and
$\lambda_{1}^{\prime}T_{i0}(\sigma\times E)^{i}$ in Hamiltonians (\ref{HNR3})
and (\ref{HNR5}), which does not reflect a direct correspondence for the
second and the third bounds of Table III, once they express constraints over
birefringent components of the tensor $(K_{F})$\ which are not embraced in
the symmetric tensor $\kappa_{\mu\nu}$. Indeed, starting from the relation
$\left(  \kappa_{DB}\right)  ^{ij}=-\left(  \kappa_{HE}\right)  ^{ij}%
=-\epsilon^{ijq}\kappa^{0q},$\ one notes that $\left(  \kappa_{DB}\right)
^{ii}=\left(  \kappa_{HE}\right)  ^{ii}=0$ (for any $i$ value),\ preventing the
association of these bounds with the symmetric $T_{\mu\nu}$\ components .

In spite of such equivalence, the tensor $T_{\mu\nu}$, as proposed in the
couplings of this letter, is not necessarily defined as symmetric and
traceless. The present approach allows to constrain pieces of the background,
as the trace of $T_{\mu\nu}$, see Eq.(\ref{Ttrace}), that would not be
restrained starting from the previous models \cite{Jonas1}, \cite{FredeNM1},
where this piece is null.

\section{New possibilities for general nonminimal couplings}

A matter of interest consists in inquiring about the possibility of other
couplings, sharing the general $CPT$-even structure of the ones already
proposed, but physically different. It is easy to notice that the couplings
(\ref{covader1}) and (\ref{covader2}) enclose the axial and non-axial options
involving two gamma matrices in the Dirac equation. A new possibility would be
associated with the following {nonminimal} derivative:
\begin{equation}
D_{\mu}=\partial_{\mu}+ieA_{\mu}+\frac{\lambda_{2}}{4}T^{\alpha\beta}%
F_{\alpha\beta}\gamma_{\mu}\gamma_{5},\label{covader3}%
\end{equation}
and with the LV Hamiltonian contribution $H_{LV}=-\lambda_{2}T^{\alpha\beta
}F_{\alpha\beta}\gamma^{0}\gamma_{5},$ which does not provide any kind of spin
interaction (neither magnetic, nor electric). Therefore, this is not an
interesting case for our purposes. The same holds for the non-axial version.
New possibilities arise when the tensors $T_{\mu\nu}$ and $F^{\alpha\beta}$
have no mutually contracted indices, leaving three free indices to be
contracted with gamma matrices, such as
\begin{equation}
D_{\mu}=\partial_{\mu}+ieA_{\mu}+i\lambda_{3}T_{\alpha\nu}F_{\mu\beta}%
\gamma^{\beta}\gamma^{\alpha}\gamma^{\nu},\label{covader4}%
\end{equation}
which implies the Lagrangian piece as%
\begin{equation}
\mathcal{L}=\bar{\Psi}(\lambda T_{\alpha\nu}F_{\mu\beta}\gamma^{\mu}%
\gamma^{\beta}\gamma^{\alpha}\gamma^{\nu})\Psi.\label{L4}%
\end{equation}
The derivative (\ref{covader4b}) is one of the combinations comprised in the
general expression
\begin{equation}
D_{\mu}=\partial_{\mu}+ieA_{\mu}+i\lambda_{3}T_{\{\alpha\nu}F_{\mu\beta
\}}\gamma^{\{\beta}\gamma^{\alpha}\gamma^{\nu\}},\label{covader4a}%
\end{equation}
with the symbol $\{\}$ denoting possible permutation of the indexes $\mu
,\nu,\alpha,\beta$. Among them, we start from the nonminimal covariant
derivative
\begin{equation}
D_{\mu}=\partial_{\mu}+ieA_{\mu}+\frac{\lambda_{3}}{8}\left(  T_{\alpha\nu
}F_{\mu\beta}+T_{\mu\beta}F_{\alpha\nu}\right)  \gamma^{\beta}\sigma^{\alpha
}{}^{\nu},\label{covader4b}%
\end{equation}
with real $\lambda_{3}$. This coupling is distinct from the all previous ones
analyzed, yielding the following Hermitian piece for the Dirac equation:
\begin{equation}
-\frac{\lambda_{3}}{8}\left(  T_{\alpha\nu}F_{\mu\beta}+T_{\mu\beta}%
F_{\alpha\nu}\right)  \sigma^{\mu\beta}\sigma^{\alpha\nu}.\label{Dirac5a}%
\end{equation}
Here only the antisymmetric part of $T_{\mu\nu}$ can contribute{\textbf{ }with
a new interaction,} due to the antisymmetry of $\sigma^{\mu\beta}$. In fact,
the symmetric part of $T_{\mu\nu}$ is associated to the Lagrangian piece,
$\lambda_{3}T^{\nu}{}_{\nu}F_{\mu\beta}\bar{\Psi}\sigma^{\mu\beta}\Psi
,$\ which provides the usual MDM interaction. The modified Dirac equation
assumes the form%

\begin{equation}
i\partial_{t}\Psi=\left[  \boldsymbol{\alpha}\cdot\boldsymbol{\pi}%
+eA_{0}+m\gamma^{0}+H_{\text{LV3}}\right]  \Psi,
\end{equation}
where the LV Hamiltonian is
\begin{equation}
H_{\text{LV3}}=-\frac{\lambda_{3}}{8}\left[  T_{\alpha\nu}F_{\mu\beta}%
^{0}+T_{\mu\beta}F_{\alpha\nu}\right]  \gamma^{0}\sigma^{\mu\beta}%
\sigma^{\alpha\nu},\label{HLV4}%
\end{equation}
whose nonrelativistic form is%
\begin{align}
&  \left.  H_{NR}=\frac{1}{2m}\left[  \boldsymbol{\pi}{}^{2}%
-e\boldsymbol{\sigma}\cdot\boldsymbol{B}\right]  +eA_{0}\right.  \nonumber\\
&  +\lambda_{3}T_{0i}E^{i}+\lambda_{3}T_{ab}\epsilon_{abc}B^{c}\nonumber\\
&  \left.  +\frac{1}{2m}\left[  (\boldsymbol{\sigma}\cdot\boldsymbol{\pi
)}H_{12}^{(3)}-H_{12}^{(3)}(\boldsymbol{\sigma}\cdot\boldsymbol{\pi})\right]
.\right.
\end{align}
where $H_{12}^{(3)}=i\lambda_{3}(T_{0i}B^{i}-T_{ab}\epsilon_{abj}E^{j}%
).$\ This coupling is not of interest for our purposes, since it does not
generate any spin interaction (at the dominant level). 

Another possibility is\textbf{ }
\begin{equation}
D_{\mu}=\partial_{\mu}+ieA_{\mu}+i\frac{\lambda_{4}}{8}\left(  T_{\alpha\nu
}F_{\mu\beta}-T_{\mu\beta}F_{\alpha\nu}\right)  \gamma^{\beta}\sigma
^{\alpha\nu},
\end{equation}
for a real $\lambda_{4}$. This coupling is also distinct from the previous
ones, leading to the Dirac equation contribution:
\begin{equation}
\frac{\lambda_{4}}{8}T^{\nu}{}_{\nu}F_{\mu\beta}\sigma^{\mu\beta}%
+i\frac{\lambda_{4}}{8}\left(  T_{\alpha\nu}F_{\mu\beta}-T_{\mu\beta}%
F_{\alpha\nu}\right)  \sigma^{\mu\beta}\sigma^{\alpha\nu}.
\end{equation}
In this case, the symmetric part of $T_{\mu\nu}$ is associated to the
Lagrangian piece, $\lambda_{4}T^{\nu}{}_{\nu}F_{\mu\beta}\bar{\Psi}\sigma
^{\mu\beta}\Psi,$\ which provides the usual MDM interaction. The modified
Dirac equation assumes the form%

\begin{equation}
i\partial_{t}\Psi=\left[  \boldsymbol{\alpha}\cdot\boldsymbol{\pi}%
+eA_{0}+m\gamma^{0}+H_{\text{LV4}}\right]  \Psi,
\end{equation}
where the LV piece is
\begin{equation}
H_{\text{LV4}}=i\frac{\lambda_{3}}{8}\left[  T_{\alpha\nu}F_{\mu\beta}%
-T_{\mu\beta}F_{\alpha\nu}\right]  \gamma^{0}\sigma^{\mu\beta}\sigma
^{\alpha\nu},
\end{equation}
and its nonrelativistic version,%
\begin{align}
&  \left.  H_{\text{NR(4)}}=\frac{1}{2m}\left[  (\boldsymbol{p}%
-e\boldsymbol{A}){}^{2}-e\boldsymbol{\sigma}\cdot\boldsymbol{B}\right]
+eA_{0}\right.  \nonumber\\
&  +\lambda_{4}T_{0i}E^{j}\epsilon_{ijk}\sigma^{k}+\lambda_{4}T_{ad}%
B^{a}\sigma^{d}\nonumber\\
&  +\frac{1}{2m}\left[  \left(  \boldsymbol{\sigma}\cdot\boldsymbol{\pi
}\right)  H_{12}^{(4)}-H_{12}^{(4)}(\boldsymbol{\sigma}\cdot\boldsymbol{\pi
})\right]  ,
\end{align}
with $H_{12}^{(4)}=i\lambda_{4}(T_{0i}\epsilon_{ikj}B^{k}\sigma^{j}%
-T_{jb}E^{j}\sigma^{b}).$ As shown in Table II, the coefficient $T_{0i}$\ is
P-odd and T-odd, being able to generate EDM, in principle. This kind of
coupling could be restrained by EDM data, using the approach of Eqs.
(\ref{Erotated1}), (\ref{bErotated1}), $\lambda_{4}\left(  T_{0i}%
\epsilon_{ijk}\sigma^{k}\right)  E^{j}=-\lambda_{4}\tilde{\sigma}^{j}E^{j},$
which yields%

\begin{equation}
\left\vert \lambda_{4}T_{0i}\right\vert \leq3.8\times10^{-16}(\mbox{GeV})^{-1}%
.\label{boundH1}%
\end{equation}
A similar analysis holds for the term $\lambda_{4}T_{ad}\sigma^{d}%
B^{a}=\lambda_{4}\hat{\sigma}^{a}B^{a},$ where $\hat{\sigma}^{a}=T_{ad}%
\sigma^{d},$ which is compatible with a non-usual MDM generation. Using MDM
data, the implied upper bound is
\begin{equation}
\left\vert \lambda_{4}T_{ad}\right\vert \leq5.5\times10^{-11}%
\,(\mbox{GeV})^{-1}.\label{boundH2}%
\end{equation}
The time-averaged sidereal variations of the bounds (\ref{boundH1}) and
(\ref{boundH2}), according to the transformation law (\ref{rotationLaw}) are,
respectively
\begin{equation}
\left\vert \lambda_{4}(-\delta_{i1}\sin\chi+\delta_{i3}\cos\chi)T_{03}%
^{\text{(Sun)}}\right\vert \leq3.8\times10^{-16}(\mbox{GeV})^{-1},
\end{equation}

\begin{equation}
\left\vert \lambda_{4}\langle\mathcal{R}_{ap}\mathcal{R}_{dq}\rangle
T_{ad}^{\text{Sun}}\right\vert \leq5.5\times10^{-11}\,(\mbox{GeV})^{-1}.
\end{equation}

One could still suppose other possibilities of coupling, similar but different
from the form (\ref{L4}). Obviously, the forms $\lambda T_{\alpha\nu}%
F_{\mu\beta}\bar{\Psi}\gamma^{\beta}\gamma^{\mu}\gamma^{\alpha}\gamma^{\nu
}\Psi$\ or $\lambda T_{\alpha\nu}F_{\mu\beta}\bar{\Psi}\gamma^{\mu}%
\gamma^{\beta}\gamma^{\nu}\gamma^{\alpha}\Psi$\ are already considered in the
Dirac term (\ref{Dirac5a}), due to the antisymmetry of $T_{\alpha\nu}%
,F_{\mu\beta}.$\ In principle, a distinct possibility could be%
\begin{equation}
\mathcal{L}=\lambda T_{\alpha\nu}F_{\mu\beta}\bar{\Psi}\gamma^{\mu}%
\gamma^{\alpha}\gamma^{\beta}\gamma^{\nu}\Psi. \label{L20}%
\end{equation}
Yet, using $\gamma^{\alpha}\gamma^{\beta}=-\gamma^{\beta}\gamma^{\alpha
}+2g^{\alpha\beta},$\ one shows that%
\begin{equation}
\mathcal{L}=-\lambda\bar{\Psi}T_{\alpha\nu}F_{\mu\beta}\gamma^{\mu}%
\gamma^{\beta}\gamma^{\alpha}\gamma^{\nu}\Psi+2\lambda\bar{\Psi}T_{\alpha\nu
}F_{\mu}^{\text{ \ }\alpha}\gamma^{\mu}\gamma^{\nu}\Psi,
\end{equation}
\ achieving the couplings of Eq. (\ref{covader4b}) and Eq. (\ref{covader2}),
with the observation that the tensor $T_{\alpha\nu}$\ now is antisymmetric.
The same holds for\ the combination $\lambda T_{\alpha\nu}F_{\mu\beta}%
\bar{\Psi}\gamma^{\mu}\gamma^{\alpha}\gamma^{\nu}\gamma^{\beta}\Psi.$\ Hence,
the Hermitian coupling corresponding to Eq. (\ref{L20}), $\lambda T_{\alpha
\nu}F_{\mu\beta}(\sigma^{\mu\alpha}\sigma^{\beta\nu}-\sigma^{\beta\nu}%
\sigma^{\mu\alpha}),$\ and others involving 4 gamma matrices, are already
contained in the previous cases.

\section{Conclusion and final remarks}

We have analysed a new class of dimension-five, $CPT$-even and
Lorentz-violating nonminimal couplings between fermions and photons, composed
of a general tensor, $T_{\mu\nu},$ in the context of the Dirac equation,
addressing its axial and non-axial versions. The nonrelativistic axial
Hamiltonian was carried out, revealing tree-level electron effective EDM and a
non-conventional MDM contribution, whereas the nonrelativistic non axial
Hamiltonian implied effective electron MDM and a non-usual EDM contribution.
The $CPT$-even nonminimal coupling proposed in Eq. (\ref{DiracM1}) evades the
Schiff's theorem, once it yields physical effects for the energy of the
system, $\Delta U=-d_{e}\cdot E,$ as explained in Ref. \cite{LeptonEDM},
allowing to directly use the electron EDM data to set upper bounds on the
nonminimal LV parameters, which does not happen for $CPT$-odd nonminimal
couplings discussed in Ref. \cite{Pospelov}. Recent experimental data about
the electron EDM and MDM were used to establish upper bounds as tight as
$\left\vert \lambda_{1}T_{00}\right\vert ,\left\vert \lambda_{1}%
T_{ii}\right\vert ,\left\vert \lambda_{1}^{\prime}T_{0i}\right\vert
,\left\vert \lambda_{4}T_{0i}\right\vert \leq3.8\times10^{-16}%
\,(\mbox{GeV})^{-1}$ and $\left\vert \lambda_{1}T_{0i}\right\vert ,\left\vert
\lambda_{4}T_{ad}\right\vert \leq5.5\times10^{-16}(\mbox{GeV})^{-1}$,
$\left\vert \lambda_{1}^{\prime}T_{ii}\right\vert \leq3.5\times10^{-11}%
\,(\mbox{GeV})^{-1},$ respectively, where the non-diagonal components bounds
involve non-conventional experiments, as explained. The sidereal analysis were
also performed. The bounds found on this axial and non-axial couplings are at
the same level of the ones obtained in Ref. \cite{Jonas1}, standing among the
best ones in the literature, if compared to bounds over dimension-five and
higher derivatives-free nonminimal LV couplings.

We point out that we have examined all the main dimension-five nonminimal
couplings involving fermions and photons, composed of a 2-rank background
tensor, the electromagnetic tensor, and gamma matrices. In accordance with our
analysis, there exist still other forms of couplings, such as%
\begin{align}
D_{\mu}  &  =\partial_{\mu}+ieA_{\mu}+\frac{\lambda_{4}^{\prime}}{2}\left(
T_{\mu\nu}F_{\text{ }\beta}^{\nu}-T_{\beta\nu}F_{\text{ \ }\mu}^{\nu}\right)
\gamma_{\alpha}\gamma^{\beta}\gamma^{\alpha},\\
D_{\mu}  &  =\partial_{\mu}+ieA_{\mu}+\frac{\lambda_{5}^{\prime}}{2}\left(
T_{\mu\nu}F_{\text{ }\beta}^{\nu}-T_{\beta\nu}F_{\text{ \ }\mu}^{\nu}\right)
\gamma_{\alpha}\gamma^{\lambda}\gamma^{\beta}\gamma^{\alpha}\gamma_{\lambda},
\end{align}
which are, however, redundant to the ones here analyzed in physical content,
being subjected to the same upper bounds [see Eqs. (\ref{bound2b}),
(\ref{bound2c}), (\ref{MDb1})].

The "rotated" MDM and EDM non-conventional interpretation could allow us to
constrain non-diagonal components of the LV tensor, as shown is Sects. II-A,
II-B and III. In order to accomplish this, it would be necessary to control the
direction of the electric or magnetic field, as well as the particle's spin
state. As an example, consider the fragment extracted from the Hamiltonian
(\ref{HNR3})\textbf{ }%
\begin{equation}
H_{\text{EDM-LV}}=\lambda_{1}(T_{00}\boldsymbol{\sigma}\cdot\boldsymbol{E}%
-T_{ij}E^{j}\sigma^{i}).\label{EDM-rotated}%
\end{equation}
If the electric field points, say, in the $\hat{\boldsymbol{x}}$\ direction,
the Hamiltonian would becomes
\begin{equation}
H_{\text{EDM-LV}}=\lambda_{1}[T_{00}\sigma_{x}E^{1}-(T_{11}E^{1}\sigma
_{x}+T_{21}E^{1}\sigma_{y}+T_{31}E^{1}\sigma_{z})],\label{EDM-rotated2}%
\end{equation}
whose expected value $\langle S|H_{\text{EDM-LV}}|S\rangle$\ depends on the
particle's spin state\textbf{. }Considering a spin polarized in the $|S_{y}%
\pm\rangle$\ state, the $T_{11}$\ and $T_{31}$\ terms average to zero,
remaining only $\left\langle S_{y}\right\rangle \neq0,$ so that following
bound can be imposed\textbf{ }%
\begin{equation}
|\lambda_{1}T_{21}|\leq3.8\times10^{-16}(\mbox{GeV})^{-1}.\label{bound3a}%
\end{equation}
A polarized beam in the $\hat{\boldsymbol{z}}$\ direction would yield the same
bound for the $T_{31}$\ term. Hence, an experimental procedure, capable of measuring spin expectation values orthogonally to the fields, could allow to constrain off-diagonal $T_{\mu\nu}$  elements using EDM and MDM data. The obtained bounds are also subject to sidereal variations,
whose transformation laws have been discussed in Sect. II-A.

\begin{acknowledgments}
The authors are grateful to CNPq, CAPES and FAPEMA (Brazilian research
agencies) for invaluable financial support. Manoel M. Ferreira Jr is grateful to: CNPq/UNIVERSAL/ 460812/2014-9; CNPq/PQ/308933/2015-0; FAPEMA/UNIVERSAL-00880/15. Rodolfo Casana is grateful  to FAPEMA/UNIVERSAL-00782/15 and CNPq/483863/2013-0.
\end{acknowledgments}

\end{document}